\documentclass[showpacs,aps,prb,amsmath,amssymb,reprint]{revtex4-1}
\usepackage{graphicx}

\begin{document}
\title{Electronic structure of fully epitaxial Co$_2$TiSn thin films}
\author{Markus Meinert}
\email{meinert@physik.uni-bielefeld.de}
\author{Jan Schmalhorst}
\author{Hendrik Wulfmeier}
\author{G\"unter Reiss}
\affiliation{Thin Films and Physics of Nanostructures, Department of Physics, Bielefeld University, 33501 Bielefeld, Germany}
\author{Elke Arenholz}
\affiliation{Advanced Light Source, Lawrence Berkeley National Laboratory, CA 94720, USA}
\author{Tanja Graf}
\author{Claudia Felser}
\affiliation{Institute of Inorganic Chemistry and Analytical Chemistry, Johannes Gutenberg University, 55128 Mainz, Germany}
\date{\today}

\begin{abstract}
In this article we report on the properties of thin films of the full Heusler compound Co$_2$TiSn prepared by DC magnetron co-sputtering. Fully epitaxial, stoichiometric films were obtained by deposition on MgO (001) substrates at substrate temperatures above 600$^\circ$C. The films are well ordered in the L2$_1$ structure, and the Curie temperature exceeds slightly the bulk value. They show a significant, isotropic magnetoresistance and the resistivity becomes strongly anomalous in the paramagnetic state. The films are weakly ferrimagnetic, with nearly 1\,$\mu_B$ on the Co atoms, and a small antiparallel Ti moment, in agreement with theoretical expectations. From comparison of x-ray absorption spectra on the Co L$_{3,2}$ edges, including circular and linear magnetic dichroism, with \textit{ab initio} calculations of the x-ray absorption and circular dichroism spectra we infer that the electronic structure of Co$_2$TiSn has essentially non-localized character. Spectral features that have not been explained in detail before, are explained here in terms of the final state band structure.

\end{abstract}

\maketitle

\section{Introduction}\label{sec:1}

The materials class of Co$_2$YZ Heusler compounds (with Y a transition metal and Z an sp element) has been the subject of extensive studies in the context of spintronics during the last decade. They are of interest because many of them are predicted as half-metallic ferromagnets with full spin polarization at the Fermi edge.

The Heusler compound Co$_2$TiSn (CTS) is of particular interest for applications. It is predicted to be a half-metallic ferromagnet with a magnetic moment of 2\,$\mu_B$/f.u. and it has a high formation energy of the Co-Ti site-swap defect.\cite{Kandpal07, Miura06} Making use of Heusler compounds which exhibit low disorder or high tolerance of the ground state properties against disorder is highly desired.

Co$_2$TiSn has been the subject of many experimental and theoretical studies. The ground state properties obtained by density functional theory (DFT) depend sensitively on the choice of the DFT method.\cite{Ishida82, Mohn95, Galanakis02, Lee05, Hickey06, Miura06, Kandpal07, Meinert10} The potential has strong non-spherical components and thus only a full-potential treatment in connection with the generalized gradient approximation (GGA) to the density yields a half-metallic ground state.\cite{Mohn95, Kandpal07}

Experiments conducted on bulk CTS find a lattice parameter of 6.07\,\AA{}, a magnetic moment of about 1.95\,$\mu_B$/f.u. and a Curie temperature ($T_C$) around 355\,K. \cite{Webster73, Kandpal07, Barth10} Further, it is found to have a strongly anomalous temperature dependence of resistivity, the temperature coefficient becomes negative above the Curie temperature. A large negative magnetoresistance reveals the importance of spin fluctuations in the compound. \cite{Majumdar05}

A rather new development aims at the magnetocaloric properties of Co$_2$TiSn, which has a large and constant Seebeck coefficient of $-50$\,$\mu$V/K above $T_C$ in the bulk.\cite{Barth10} There have been some efforts to understand the unusual transport properties of CTS by \textit{ab initio} band structure and semi-classical transport theory.\cite{Barth10, Sharma10} These properties make CTS interesting for a possible application in spin caloritronics, which attempt to make use of the interactions between heat and spin. An implementation into thin films is of particular importance for such applications.

Only two studies on thin films of CTS are available as far as we know. Gupta \textit{et al.} applied pulsed laser ablation to grow CTS on Si (001) substrates from a stoichiometric target at growth temperatures up to 200$^\circ$C.\cite{Gupta09} The authors found off-stoichiometric, polycrystalline films with (022) texture. Suharyadi \textit{et al.} utilized an atomically controlled alternate deposition technique based on electron beam evaporation.\cite{Suharyadi10} They have grown (001) oriented, L2$_1$ ordered films on Cr buffered MgO (001) substrates at growth temperatures up to 600$^\circ$C and investigated them by nuclear resonant scattering.

In this paper we present a successful preparation technique based on DC magnetron co-sputtering. We present data on the structural and magnetic properties of our films. Further, we characterize the electronic transport properties which make CTS a particularly interesting compound. Finally we discuss the electronic structure of our CTS films based on soft x-ray absorption spectroscopy and \textit{ab initio} electronic structure calculations.

\subsection{Experimental details}

The samples were deposited using a UHV sputtering system equipped with five DC and two RF 3" magnetron sputtering sources arranged in a confocal sputter-up geometry. Material can be deposited from up to four sources simultaneously. The target-to-substrate distance is 21\,cm and the inclination of the sources is 30$^\circ$. The substrate can be rotated at 30\,rpm and heated to 1000$^\circ$C by a ceramic heater via radiation from the substrate carrier backside. The deposition process can be controlled with an \textit{in situ} quartz film thickness monitor, which can be moved to the position, where the substrate is located during deposition. Moreover, an electron beam evaporator with one crucible is placed in the center of the chamber. Film thickness homogeneity is better than $\pm$ 5\,\% over a substrate diameter of 4" for sputtering as well as evaporation with rotation turned on. The base pressure of the system is typically better than $3 \cdot 10^{-9}$\,mbar.

Using the quartz sensor and x-ray reflectometry (XRR), the film stoichiometry of a compound can be set up with a relative accuracy of about $\pm 10\,\%$. Using inductively coupled plasma optical emission spectroscopy (ICP-OES) the sputter parameters were fine-tuned. For the samples deposited at high temperature we checked the stoichiometry by energy dispersive x-ray analysis (EDX) in an electron microscope and found no deviation from the stoichiometry of room temperature deposited films of same thickness. The sputtering power ratios were 1\,:\,1.67\,:\,0.34 (Co:Ti:Sn). The voltages were constantly monitored during the deposition, which remained constant throughout all deposition processes, ensuring the reproducibility of the method. Cross-talk effects on the Sn target constituted a serious problem for the deposition process due to the low sputtering power applied to the source. This was suppressed by a chimney-like cylinder put around the source, such that there was no line-of-sight from this target to another. The compound was deposited at a rate of 1.5\,\AA/s. Sample rotation was set to 28 rpm, making sure that with each turn only one primitive cell was deposited. All elemental targets had 4N purity. The sputtering pressure was set to $2\cdot10^{-3}$ mbar. With this technique we have fabricated thin film samples with a precisely set up stoichiometry of Co$_{2.0}$Ti$_{1.0}$Sn$_{1.0}$, with errors of $< 3\%$ for the individual constituents. 

All samples used in this study had the following stack sequence: MgO (001) / MgO 5\,nm / CTS 18\,nm / MgO 2\,nm. The lower MgO was deposited by RF sputtering at $2.3\cdot 10^{-2}$\,mbar to ensure good crystallinity of the buffer. The upper MgO was deposited by e-beam evaporation from single crystal MgO slabs after cooling the samples to less than 100$^\circ$C. The base pressure during deposition with the heated substrate was always below $5\cdot 10^{-8}$\,mbar. 

Resistivity was measured in standard in-line four-probe DC geometry in a closed-cycle He cryostat and a vacuum furnace. The resistivity $\rho$ is calculated from the film thickness $d$, the voltage $U$ and the current $I$ as $\rho = d \cdot (\pi / \ln{2}) \cdot (U/I)$. Magnetoresistance was measured using variable permanent magnet (coaxial Halbach cylinder configuration, Magnetic Solutions Multimag) with a maximum field strength of 10\,kOe in the cryostat. The data were taken by driving full magnetization loops and then averaging the points for each field magnitude.

The Seebeck coefficient was determined in a home built setup in air. The sample was contacted with platinum tips. It was measured at an average temperature of $\bar{T}=310$\,K with a temperature gradient of $\Delta T = 10$\,K.

Magnetic measurements were taken using a superconducting quantum interference device (SQUID) at temperatures in the range of 5\,K to 400\,K in magnetic fields of up to 50\,kOe. 

X-ray diffraction (XRD) and reflectometry (XRR) have been performed using a Philips X'Pert Pro MPD with a Cu source in Bragg-Brentano configuration. Texture characterization was additionally performed with collimator point focus optics on an open Eulerian cradle.

Temperature dependent x-ray absorption spectroscopy (XAS), x-ray magnetic circular dichroism (XMCD) and x-ray magnetic linear dichroism (XMLD) was performed at BL 6.3.1 and BL 4.0.2 of the Advanced Light Source in Berkeley, USA. The element-specific magnetic properties  were investigated at the Co- and Ti-$L_{3,2}$ edges in surface-sensitive total electron yield mode (TEY)\cite{Idzerda94} for temperatures between 20K and 370K.
For XMCD, the sample was saturated by applying a magnetic field of max. $\pm$ 20\,kOe along the x-ray beam direction using elliptically polarized radiation with a  polarization of $P_{h\nu}=\pm60\%$ (BL 6.3.1) and $P_{h\nu}=\pm90\%$ (BL 4.0.2), respectively. The x-rays angle  of incidence with respect to the sample surface was $\alpha = 30^\circ$ (BL 6.3.1) and $\alpha = 90^\circ$ (BL 4.0.2), respectively. $I^+$ and $I^-$ denote the absorption spectra (normalized to the x-ray flux measured by the total electron yield of a Au grid in front of the sample) for parallel and anti-parallel orientation of the  photon spin and the magnetization of the sample.   
The XAS and XMCD spectra are defined as XAS$^{c}=(I^+ + I^-)/2$ and XMCD$=(I^+-I^-)$, respectively. To calculate the element-specific spin and orbital magnetic moments from the data we applied sum-rule analysis;\cite{Chen95} details of the procedure are presented in the Appendix. 

Anisotropic XMLD spectra were taken at BL 4.0.2 with 100\% linearly polarized light in normal incidence using the eight-pole electromagnet end station.\cite{Arenholz05} The magnetic field for switching the magnetization of the sample was applied parallel and orthogonal to the polarization vector of the incoming light, the according absorption spectra normalized to the x-ray flux are denoted as $I^{\parallel }$ and $I^{\perp }$. The XAS and XMLD spectra are then defined as XAS$^{l}=(I^{\parallel }+I^{\perp })/2$ and XMLD$=(I^{\parallel }-I^{\perp })$, respectively. Spectra were taken with magnetic fields aligned along the [100] and the [110] directions of the Co$_2$TiSn films. The applied magnetic field of 4.5\,kOe was canted out of the surface plane by 10$^\circ$ to improve the electron yield signal. However, the XMLD results are nearly unaffected by this because the demagnetizing field perpendicular to the film plane is so strong that the magnetization is tilted out-of-plane by less than $5^\circ$ (measured by analyzing the XMCD asymmetry for different tilting angles).

The XMCD and XMLD spectra were taken by switching the magnetic field at each energy point. To remove non-dichroic artifacts we performed measurements for positive and negative polarization (XMCD) or different spatial orientations of the polarization vector (XMLD) and averaged the corresponding spectra.

\subsection{Theoretical approach}

\begin{table}[b]
\caption{Results of band structure calculations with SPRKKR. The magnetic moments are given in $\mu_B$\,/\,atom.}
\begin{ruledtabular}
\begin{tabular}{ c c c c c c }
$m_\text{spin}^\text{Co}$ & $m_\text{orb}^\text{Co}$ & $N_h^\text{Co}$ & $m_\text{spin}^\text{Ti}$ & $m_\text{orb}^\text{Ti}$ & $N_h^\text{Ti}$ \\ \hline
0.96	& 0.04	& 2.06	& -0.03	& 0.01	& 7.65
\end{tabular}
\end{ruledtabular}
\label{Tab:SPRKKR}
\end{table}

The electronic structure probed by x-ray absorption spectroscopy has been investigated in direct comparison with \textit{ab initio} electronic structure calculations. We used two different approaches to this end. First, electronic structure calculations were performed with the \textit{Munich} SPRKKR package, a spin-polarized relativistic Korringa-Kohn-Rostoker method.\cite{sprkkr} And second, in order to take care of the excited state band structure, which is actually probed in XAS, spectrum simulations were carried out in FEFF9, a relativistic real-space full multiple scattering code.\cite{feff9}

In SPRKKR, the band structure and the ground state properties were calculated in the fully-relativistic representation of the valence states, thus including spin-orbit coupling. The angular momentum cutoff was set to $l_\text{max} = 3$ (spdf-basis) and the full potential was taken into account. The bulk lattice parameter of $a = 6.07$\,\AA{} was used. The exchange-correlation potential was modeled by the generalized gradient approximation (GGA) in the Perdew-Burke-Ernzerhof parametrization.

The resulting atomic magnetic moments were then used as input parameters to FEFF9, which is not spin self-consistent. The self-consistent potential was obtained on a cluster of 59 atoms and the x-ray absorption near edge spectrosopy (XANES) was calculated on a cluster of 229 atoms. The complex Hedin-Lundqvist self-energy was applied and the calculations were done with the final state rule, including a full screened core hole on the absorber. The angular momentum for the full multiple scattering was taken to $l_\text{max} = 3$.

The ground state described by the SPRKKR calculation is not half-metallic with the experimental lattice parameter, in contrast to calculations with full potential linearized augmented plane-waves codes. \cite{Kandpal07, Meinert10} The Fermi energy is slightly above the minority spin gap; a small increase of the lattice parameter would move $E_F$ into the gap. However, this does not significantly change the shape of the calculated XAS spectrum. The total spin moment is 1.9$\,\mu_B$\,/\,f.u. and the total orbital moment is 0.09$\,\mu_B$\,/\,f.u.. The atom-resolved magnetic moments and the numbers of holes for Co and Ti are given in Tab. \ref{Tab:SPRKKR}. The negative Ti spin moment indicates a weakly ferrimagnetic behavior of CTS.

\section{Experimental results}

\subsection{Structure}\label{sec:2}

\begin{figure}[t]
\includegraphics[width=8cm]{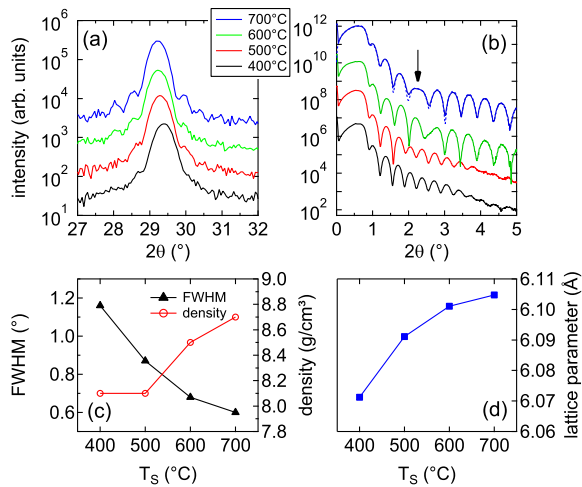}
\caption{(a): X-ray diffraction scans of the (002) reflections showing Laue oscillations. (b) X-ray reflectometry (XRR) scans. The dashed line represents the best fit to the experimental curve with $T_S = 700^\circ$C. (c): Full-widths at half-maximum (FWHM) of the rocking curves and effective density determined by XRR. (d): Out-of-plane lattice parameter $c$. }
\label{FigXRDLayout}
\end{figure}

XRD and XRR were utilized to investigate the structure of the films. Figure \ref{FigXRDLayout} displays a set of data that were extracted from the measurements. As is clearly visible in Figure \ref{FigXRDLayout} (a), the films show Laue oscillations on the (002) reflection that become more pronounced with increasing deposition temperature. Laue oscillations are only observed if the crystalline coherence is very good and the roughness is small. While the two films deposited at lower temperatures show only weak oscillations, the two films deposited at higher temperature exhibit pronounced fringes. Only weak asymmetry of the fringes is observed for $T_S = 700^\circ$C, indicating nearly homogeneous (or no) strain along the growth direction.

Four intense (111) reflections have been observed in pole figure analysis at the expected tilt angle of $\Psi = 54.74^\circ$ for all samples. The intensity increases considerably with increasing deposition temperature. The epitaxial relationship is Co$_2$TiSn [100] $\parallel$ MgO [110], which is commonly observed for Heusler compounds on MgO (001) substrates.

The out-of-plane lattice parameter $c$ measured on the (004) reflection, displayed in Figure \ref{FigXRDLayout} (d), is found to increase with increasing deposition temperature and converges for the highest deposition temperatures. For 700$^\circ$C, we find a lattice parameter of $c=6.105$\,\AA{}.

The full-width at half-maximum (FWHM) of the rocking curves measured on the (004) reflections are displayed together with the density determined by XRR in Figure \ref{FigXRDLayout} (c). For high deposition temperature the rocking curve FWHM is found to be as low as 0.6$^\circ$, which demonstrates the narrow orientation distribution of the individual film grains.

XRR provides indirect information on the film morphology. The density determined by XRR has to be seen as an effective density, which only reflects the real film density if the surface roughness is low with a Gaussian distribution on a small lateral scale. In Fig. \ref{FigXRDLayout} (b) we present the XRR curves of our samples. The roughness is high for the two samples with lower growth temperature, which is identified by a quick vanishing of the Kiessing fringes. The MgO cover layer does not show up as an individual resonance. We find an increase in the XRR density for deposition with 600$^\circ$C and more, while the roughness is greatly reduced and the MgO cover layer becomes visible (see arrow). The XRR roughness of the film with $T_S = 700^\circ$C is 0.3\,nm. The scans for the two lower deposition temperatures can not be fit with the Parratt algorithm.\cite{Parratt54} They show two main Fourier components at 18\,nm and 23\,nm, and a difference component at 5\,nm. A columnar growth with high and low grains that have $18 \pm 5$\,nm thickness can be inferred from this. At higher temperatures, the growth changes to a mode with large and smooth grains of equal height. This behavior has been confirmed by atomic force microscopy.

From Thornton's model\cite{Thornton74} of film growth for sputtered films it is expected to find a transition from a fine-grained columnar structure to a regime with large grains governed by bulk diffusion and recrystallization at about half the melting temperature, $T_S / T_m \approx 0.5$. In fact, the melting point of Co$_2$TiSn is 1720(20)\,K,\cite{Balke_private} i.e., this transition is expected around 600$^\circ$C. 

Using the experimental bulk lattice parameter $a=6.07$\AA{}, the density of the compound is calculated to be 8.446 g/cm$^3$. If one assumes a perfect, strained epitaxial growth on the MgO substrate, the lattice will be distorted tetragonally, with an in-plane lattice parameter $a = \sqrt{2} \cdot 4.21 \text{\AA{}} = 5.95 \text{\AA{}}$ and accordingly expanded out-of-plane. If the volume remained constant, the out-of-plane lattice parameter would be 6.32\,\AA{}. For the film deposited at 700$^\circ$C, we measured $c = 6.105$\,\AA{}. Recalculating the density for this tetragonal configuration gives $\rho = 8.74$\, g/cm$^3$, which is in close agreement with the measured density of $\rho$ = 8.7 g/cm$^3$. This result supports the growth model discussed above. Further, we have shown in a recent paper by \textit{ab initio} theory that a tetragonal distortion of Co$_2$TiSn can easily occur because of the low energy associated with the distortion.\cite{Meinert10} It is of the order of 50\,meV / f.u., and is thus easily activated during the growth. However, at lower temperatures this constitutes a metastable state.

A detailed analysis of the chemical order in the films was performed using anomalous x-ray diffraction. The results from this study will be published elsewhere in detail. The main result is however, that the atomic disorder decreases considerably with increasing deposition temperature and that the films deposited at 700$^\circ$C have a high degree of L2$_1$ order.

\subsection{Magnetism}

\begin{figure}[b]
\includegraphics[scale=1]{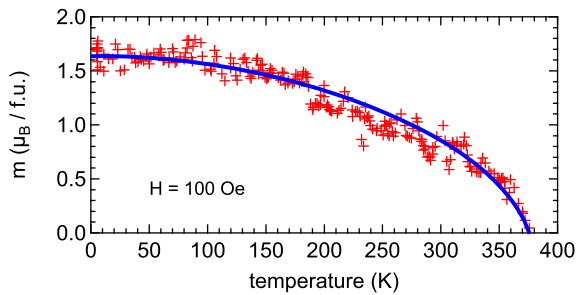}
\caption{Magnetization in dependence of the sample temperature (markers).  It was taken as a temperature sweep with a constant field of 100\,Oe. The solid line is a guide to the eye.}
\label{fig:squid}
\end{figure}

Measurements taken on the sample with $T_S = 700^\circ$C give a magnetic moment of $m = 1.6(1)\,\mu_B$\,/\,f.u. and a Curie temperature of $T_C = 375(5)$\,K (Fig. \ref{fig:squid}). The Curie temperature is higher than in bulk samples, where it has been reported to be about 355\,K. Our value is in better agreement with a first principles calculation based on the mean-field approximation, which gave 386\,K.\cite{Meinert_PRB_submitted} The coercive field is 160\,Oe at 20\,K and 150\,Oe at room temperature. Since the magnetization declines sharply at $T_C$,  we can conclude that the films consist of a single magnetic phase.

\subsection{Electronic transport}

\begin{figure}[t]
\begin{center}
\includegraphics[scale=1]{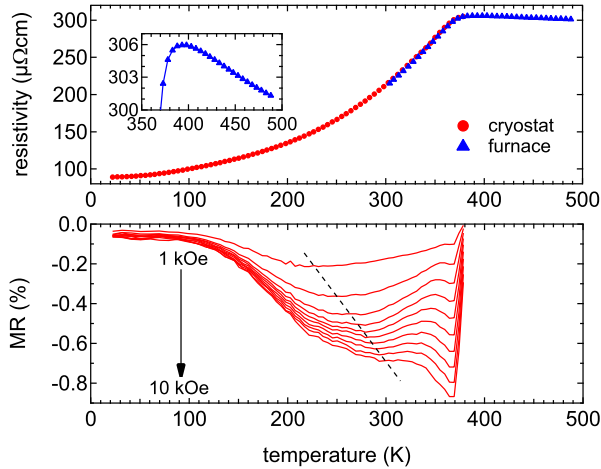}
\caption{Top: Resistivity of a Co$_2$TiSn film deposited at $T_S=700^\circ$C on a MgO single crystal. The inset shows the region around the ferromagnet-paramagnet transition. Bottom: Corresponding magnetoresistance for fields of 1\,kOe to 10\,kOe with the magnetic field $\vec{H}$ in the sample plane and the current $\vec{j} \perp \vec{H}$.}
\label{fig:resistivity}
\end{center}
\end{figure}

Resistivity and magnetoresistance have been measured on a sample deposited at $T_D = 700^\circ$C; the data are shown in Fig. \ref{fig:resistivity}. The resistivity shows clearly the cusp-type resistivity anomaly that is also observed for bulk samples of Co$_2$TiSn at $T_C$. Details of the transition can be found by analyzing the first and second derivatives of the resistivity curve.

The onset of the transition is defined as the first inflection point of the resistivity; it is found at 350(5)\,K. The maximum of the resistivity is at 395(5)\, K, i.e., 20\,K above $T_C$. The offset of the transition, given by the second inflection point, is at 440(5)\,K. At $T_C = 375(5)$\,K we find the maximal change rate of the resistivity's slope, identified by a clear minimum of the second derivative.

By plotting the logarithm of the resistivity against $1/T$ for the data points above the second inflection point, we find the effective gap width of the paramagnetic state to be $E_g = 6.5 \pm 0.5$\,meV. This is considerably smaller than the gap width of $12.7\pm1$\,meV  reported for bulk samples. However, it has been argued by Barth \textit{et al.} that an actual transition to a semiconductor is improbable. They found significant differences for the calculated conductivity tensors between spin-polarized and unpolarized calculations. By mixing the states weighted by a molecular field approximation for the magnetization, they could partly explain the anomalous behavior of the resistivity.\cite{Barth10} 

Compared with bulk samples, we also find a notably lower residual resistivity $\rho(\text{20K}) = 89\,\mu\Omega$ cm and a total resistivity amplitude $ ( \rho_\text{max} - \rho_\text{min} ) = 216\,\mu\Omega$cm, compared to 310 and $205\,\mu\Omega$cm,\cite{Majumdar05} or 245 and 135\,$\mu\Omega$cm,\cite{Barth10} respectively. The residual resistivity of a metal is mainly given by its defect density, i.e., dislocations, disorder, impurities and grain boundaries. In a thin film, one has to take the interfacial scattering into account. Our thin films have very low residual resistivity compared to bulk samples, which might indicate that their crystalline properties are superior to those of bulk samples. We attribute this to large, flat grains and good chemical order.

The temperature dependence of the resistivity is well described by a $T^2$ term up to 180\,K, which is mainly attributed to electron-electron scattering. Above 180\,K up to the first inflection point the curve is better fit by a $T^3$ law. In bulk samples, the parabolic shape of the resistivity curve at intermediate temperatures is less pronounced than in our films. However, the overall shape is in agreement with the curves found by other authors.

The magnetoresistance (MR) of the film, defined by $\text{MR}(H,T) = (\rho(H,T) - \rho(0,T))/\rho(0,T)$,  shows strongly nonlinear behavior. At low temperature only weak MR is found. With increasing temperature an increasing MR is observed, which is negative over the whole temperature range, i.e., the resistivity is lower if a magnetic field is applied. It has a pronounced, nonlinear dependence on the applied magnetic field. With an available magnetic field of 10\,kOe the MR was by far not saturated. A distinct extremum is observed at large fields right below $T_C$, being the global minimum of the curve at fields larger than 7\,kOe. Above $T_C$ the MR vanishes. The appearance of the extremum and its amplitude are in agreement with the data published by Majumdar \textit{et al.}.\cite{Majumdar05} The MR can be explained in terms of spin fluctuations and associated spin-flip scattering: at low temperature, the fluctuations are nearly zero and a small magnetic field is sufficient to saturate the film. With increasing temperature, fluctuations become more important, but can be suppressed by enforcing a particular spin orientation in a strong field. This picture is supported by the shift of the first minimum with increasing magnetic field, denoted by the dashed line in Fig. \ref{fig:resistivity}. The MR is enhanced at $T_C$ because the spin fluctuations are strongest at the transition temperature and the ferromagnetic state is stabilized in a large field. Furthermore, the MR has no traceable anisotropic MR (AMR) contribution: the typical inversion of the MR at zero field for $\vec{j} \perp \vec{H}$ compared to $\vec{j} \parallel \vec{H}$ is missing.

The Seebeck effect has been measured on the same sample as the resistivity. It was $S = -14 \pm 2\,\mu$V\,/\,K at 310\,K, which is about 2.6 times lower than in the bulk ($-37 \pm 2\,\mu$V\,/\,K)\cite{Barth10}. This is in agreement with the much lower resistivity of our films compared to bulk samples. Barth \textit{et al.} point out that the Seebeck coefficient can be enhanced by scattering on grain boundaries or impurities, which appear to be rarer in the films. On the other hand, the Seebeck coefficient is proportional to $\nu / \sigma$, with the electrical conductivity $\sigma$ and the thermal conductivity $\nu$. Thus, the lower $S$ may also indicate a lower heat conductivity of the film.

\subsection{Interfacial chemistry}

XMCD and XAS$^c$ measurements were performed at BL 6.3.1 at 20K and at RT for films deposited on MgO single crystalline substrates (T$_S$=400$^\circ$C, 500$^\circ$C, 600$^\circ$C, 700$^\circ$C and post-annealed samples).  

The Co XMCD signals for different deposition temperatures show two notable trends: the Co magnetic moment, measured at 20\,K, and the ratio of the Co XMCD signals measured at RT and at 20K increase with increasing T$_S$. This implies that the chemical order improves with increasing substrate temperature, resulting in higher saturation magnetization and higher Curie temperature. That is in agreement with SQUID measurements on the same samples.

At $T_S = 400, 500^\circ$C we found multiplet structures on the Ti L$_{3,2}$ edges, which indicate formation of interfacial TiO$_2$. \cite{Brydson89} These structures almost vanish at $T_S = 600^\circ$C and are not traceable at $T_S = 700^\circ$C anymore. The spectral shapes of the XMCD signals on Co and Ti do not change on the other hand, only the amplitude is reduced at lower deposition temperature. The large roughness of the films deposited at the lower temperatures leads to an incomplete covering with the protective MgO layer. The CTS compound is thus oxidized in air, which is particularly observed as surfacial TiO$_2$, which is not magnetic.

In vacuum post-annealed samples have been additionally investigated for their interfacial chemistry. Annealing at temperatures above 350$^\circ$C resulted in formation of interfacial TiO$_2$. Naturally, this will also happen at the lower interface to the MgO substrate. Because of the high growth temperatures, we can expect an oxide thickness of several nanometers. This effect may account for the low average magnetization measured in the SQUID. An oxidized bottom layer of 3\,nm thickness can account for the deviation from the nearly 2\,$\mu_B$\,/\,f.u. measured in the bulk and predicted theoretically.

Using the results from this systematic analysis we chose two samples for in-detail investigations described in the next section.

\subsection{Element specific magnetization}

\begin{figure}[t]
\begin{center}
\includegraphics[width=8cm]{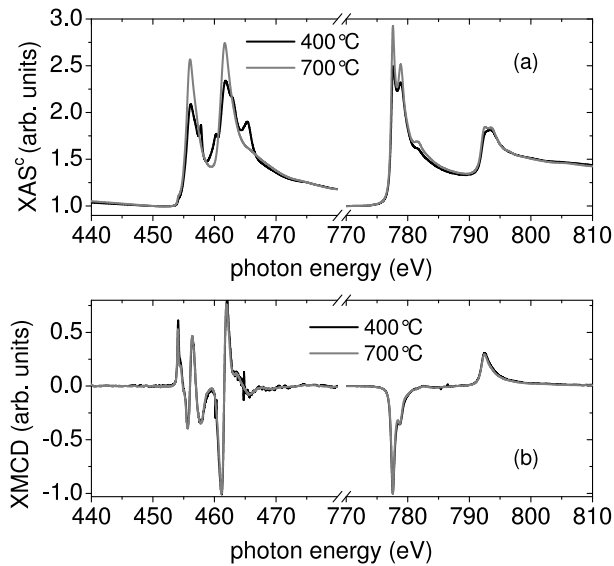}
\caption{Normalized XAS$^c$ and XMCD spectra of Ti and Co measured at 20K for samples deposited on MgO single crystals at 400$^\circ$C and 700$^\circ$C, respectively.}
\label{fig:C}
\end{center}
\end{figure}

\begin{figure}[t]
\begin{center}
\includegraphics[width=8cm]{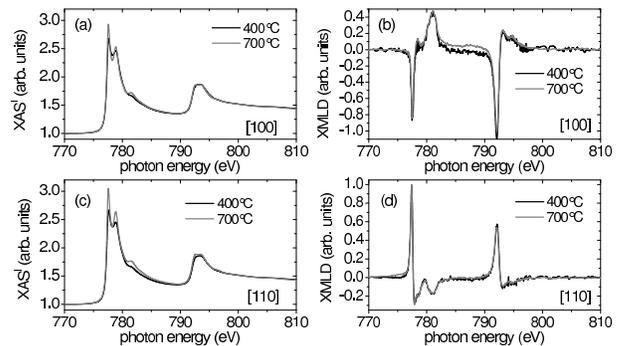}
\caption{Normalized XAS$^l$ and XMLD spectra of Co measured at 20K in the [100] (a and b) and [110] (c and d) directions for samples deposited on MgO single crystals at 400$^\circ$C and 700$^\circ$C, respectively.}
\label{fig:D}
\end{center}
\end{figure}

Highly resolved XMCD and XMLD spectra were taken at BL 4.0.2 at 20K for the samples deposited at 400$^\circ$C and 700$^\circ$C, respectively (see Fig. \ref{fig:C} and \ref{fig:D}). 

Whereas the XAS$^c$ spectra show significant differences for the two deposition temperatures for Co and Ti, the shape of the XMCD spectra does not depend on the deposition conditions. For Co the deposition at higher temperature results in a more pronounced fine structure, consisting of a double peak at the L$_3$ resonance and a shoulder about 4\,eV above the resonance. These structures are also reflected in the L$_2$ resonance, but less pronounced. Klaer \textit{et al.} investigated Co$_2$TiSn bulk samples ({\it in situ} fractured in UHV for XMCD investigation).\cite{Klaer09}  They also observed a double peak structure at the L$_3$ resonance, but less pronounced compared to our sample deposited at 700$^\circ$C. Moreover, the double peak structure at the L$_2$ edge was not found in these bulk samples. Yamasaki {\it et al.} \cite{Yamasaki02} have also investigated bulk samples ({\it in situ} scraped in vacuum for XMCD investigation), but in contrast to the results by Klaer \textit{et al.} and us they observed three separated peaks at the L$_3$ edge and only one broad peak at the L$_2$ resonance. Obviously, their samples had a different electronic structure.

Our Co XMCD spectra also show the double peak structure at the L$_3$ edge, while at the L$_2$ edge only a shoulder is visible. Again, the structures in our XMCD spectra are sharper than those given by Klaer \textit{et al.} and Yamasaki \textit{et al.}. Our Ti XMCD spectra shown in Fig. \ref{fig:C}b are similar  to the data by Klaer {\it et al.}; Yamasaki {\it et al.} do not provide data on the Ti L-edges. However, the shape is very different compared to data collected by Scherz {\it et al.} \cite{Scherz04} on the system Fe/Ti/Fe(110). Therefore the relative alignment of the Co and Ti magnetic moments is not obvious from a comparison with their reference data.

In order to get further insight into the element specific magnetic properties, we applied the XMCD sum rules, which give information on the magnitude and sign of the element specific spin and orbital moments and the number of d-holes. Details of the procedure can be found in the Appendix. The results of the sum-rule analysis for the Co XMCD spectra are summarized in Tab. \ref{Tab:XMCD1}. 

\begin{table}[b]
\caption{Results of the sum rule analysis of the Co XMCD spectra measured at 20K for the samples deposited at 400$^\circ$C and 700$^\circ$C, respectively. }
\begin{ruledtabular}
\begin{tabular}{ c c c c c }

T$_S$ & $m_{spin}$ & $m_{orb}$ &  $m_{orb} / m_{spin}$    &  $N_h$  \\ \hline
 400$^\circ$C& $0.48\mu_B$ &   $0.025\mu_B$   & 5.2\%  &  1.50 \\ 
 700$^\circ$C&$0.98\mu_B$&   $0.055\mu_B$ & 5.6\%   &  1.75  \\
\end{tabular}
\end{ruledtabular}
\label{Tab:XMCD1}
\end{table}

The Co spin moment is close to 1$\mu_B$ for a deposition temperature of 700$^\circ$C. For the deposition at 400$^\circ$C the Co spin moment is a factor of two smaller, but the orbital to spin moment ratio is nearly identical for both deposition temperatures; the orbital moment is parallel to the spin moment. Both the spin and orbital moments are in very good agreement with the theoretical results. The number of d-holes is lower than for pure Co metal (1.75 and 1.5 for Co$_2$TiSn deposited at 700$^\circ$C and 400$^\circ$C, respectively,  and 2.4 for pure Co\cite{ScherzPhD}), which indicates a rather large charge transfer to the Co d states in Co$_2$TiSn. It is actually even a bit lower than the theoretical value of 2.06.

While the sum rules work well for Co, dynamical screening effects of the x-ray field prohibit their direct application to the early 3d transition metals.\cite{Ankudinov03} This core-hole - photoelectron interaction leads to an intermixing of the L$_3$ and L$_2$ resonances, which is the reason for the deviation from the statistical branching ratio of 2:1 for the two edges. The intermixing, also known as jj-mixing of the 2p$_{1/2}$ and 2p$_{3/2}$ levels,  leads to wrong results when the sum rules are applied to the early 3d transition metals. It has been suggested by Scherz that one can estimate the Ti spin moment by multiplying the result from the sum rule analysis by a factor of 4.\cite{ScherzPhD} This result has been obtained on the Fe/Ti/Fe(110) trilayer system. On the other hand, it must be expected that this correction factor itself depends on the actual electronic structure and thus the screening strength. The direct result from the sum rule analysis is $m_{spin}=-0.038\,\mu_B$ for the sample deposited at 700$^\circ$C, which is in good agreement with the theoretical result. In particular, an anti-parallel alignment with the Co spin moment is found.
It is worth to mention, that the Ti orbital moment (the apparent value is  $m_{orb}=0.022\,\mu_B$) is aligned anti-parallel to the Ti spin moment. The latter is in accordance with Hund's rules, which expect an anti-parallel alignment of the spin and orbital moment, because the Ti 3d shell is less than half filled. Because of the formation of interfacial TiO$_2$ the XMCD data can not be quantified for $T_S=400^\circ$C. However, all qualitative conclusions with respect to the alignment of the Co and Ti orbital and spin moments are preserved for lower deposition temperatures, because the shapes of the Co and Ti XMCD spectra do not depend on $T_S$. 
In summary, the XMCD results are in very good agreement with theoretical expectations.

\begin{figure}
\begin{center}
\includegraphics[width=8cm]{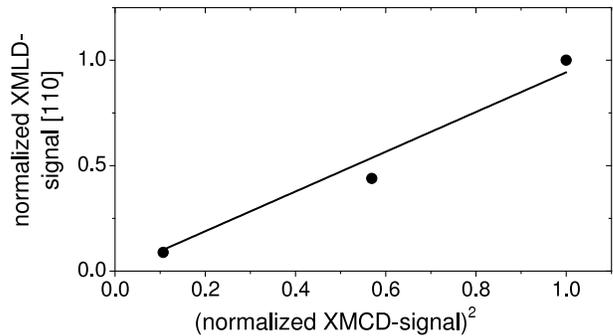}
\caption{Normalized Co XMLD signal for the [110] direction as a function of the square of the normalized XMCD signal. The data points correspond to measurements taken at 20K, 300K and 370K. The sample was deposited on MgO single crystals at 700$^\circ$C.}
\label{fig:B}
\end{center}
\end{figure}

In general it is expected, that the XMLD signal is proportional to the square of the total magnetic moment of the individual atoms ($\text{XMLD}=\beta_l \cdot \left<m_\text{total}\right>^2$), whereas the XMCD signal should be directly proportional to the magnetic moment ($\text{XMCD}=\beta_c \cdot \left<m_\text{total}\right>$).\cite{Kunes03} Comparing the XMCD and XMLD signals (normalized to the post-edge jump height $\eta$, because the number of 3d-holes $N_h$ is different for the samples deposited at 400$^\circ$C and 700$^\circ$C, respectively) for Co, it is interesting to note that XMLD/XMCD$^2$ is about 65\% larger for the sample deposited at 400$^\circ$C than for the 700$^\circ$C sample. In the simple picture that the proportionality factors $\beta_c$ and $\beta_l$ are the same for both deposition temperatures, this means that in the disordered 400$^\circ$C sample some of the Co atoms are anti-ferromagnetically coupled to the other Co atoms. On the other hand it is known, that the XMLD effect can become quite large in systems with localized electrons. The magnitude of the XMLD is given essentially by the magnetic moment and the 2p level exchange splitting, which itself is proportional to the magnetic moment. Actually, without the exchange splitting of the 2p levels, the XMLD would vanish. Localized 3d electron states increase the 2p-3d exchange interaction, giving rise to an enhanced XMLD. \cite{Freeman06} Therefore, the decrease of XMLD/XMCD$^2$ with the deposition temperature could also hint to a higher degree of localization of the Co moments for the 400$^\circ$C sample. This is in agreement with an oxidized surface, in which the electrons should be more localized. However, the fine structure at the Co-$L$ edges becomes more pronounced for higher deposition temperature (see Fig. \ref{fig:C}a, \ref{fig:D}a and \ref{fig:D}c) which might indicate a higher degree of localization for higher deposition temperatures. The electron localization would give the Co a more atomic character, and atomic multiplets would become important, giving rise to a fine structure on the x-ray absorption spectrum. On the other hand, this would contradict the XMLD result. The maximum amplitude of the XMLD for $T_S = 700^\circ$C is 5.7\,\% at the Co L$_3$ edge in the [110] direction. Thus, the Co 3d states take an intermediate position between the elemental ferromagnets Co and Fe, that have around 2\,\%, and strongly localized systems like Mn in (Ga, Mn)As, which has about 12\,\%.\cite{Freeman06} Obviously, this discrepancy needs to be investigated by direct \textit{ab initio} calculations of the absorption spectra, which will be discussed in Sec. \ref{sec:3}. 

For the sample deposited at 700$^\circ$C the XMCD and XMLD effect was studied also at elevated temperatures. The normalized XMCD signals of Co and Ti have the same shape at 20\,K, 300\,K and 370\,K. Furthermore, the Ti XMCD asymmetry changes by the same factor as the Co asymmetry between 20K and 370K. Therefore the ratio between the Ti and Co magnetic moments is not significantly changed at elevated temperatures.

The temperature dependence of the XMLD signal was measured for the [110] direction. As shown in Fig. \ref{fig:B} the XMLD signal scales well with XMCD$^2$, which was also found for other materials like (Ga,Mn)As \cite{Freeman06} in accordance with the above mentioned expectation.

\section{Electronic structure}\label{sec:3}

\begin{figure}
\begin{center}
\includegraphics[width=8cm]{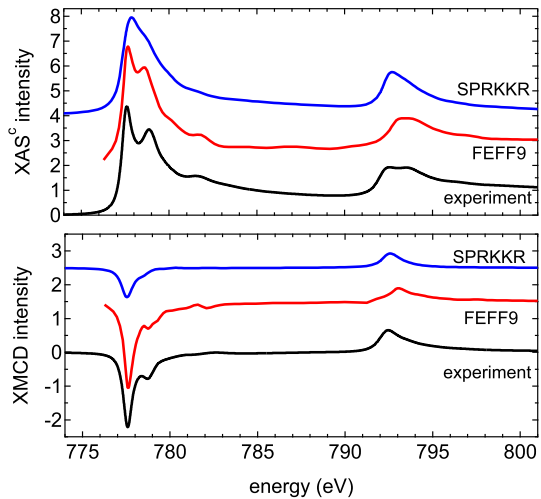}
\caption{Comparison of the calculated Co L$_{3,2}$ XAS and XMCD spectra carried out in FEFF9 and SPRKKR to experimental spectra. The XMCD signals have been scaled to 90\% to account for the experimental polarization degree. The experimental and the FEFF9 spectra are scaled to 1 in the post-edge region. The SPRKKR spectra are scaled to match the experimental L$_3$ resonance. The theoretical spectra are aligned in energy with the experimental spectrum.}
\label{fig:Co_comparison}
\end{center}
\end{figure}

As discussed above, the fine structure observed at the Co L$_{3,2}$ edges can have its origin in atomic multiplet effects related to electron localization or simply in the particular (itinerant) electronic structure of Co$_2$TiSn. The experimental XAS and XMCD spectra are compared to calculations with SPRKKR and FEFF9 in Fig. \ref{fig:Co_comparison}.

The SPRKKR spectra show broad edges and some weak shoulders on the high energy side of the white lines. The XMCD intensity is significantly too small with respect to the XAS intensity. Further, the ratio of the L$_3$ and L$_2$ XMCD signals is incorrect, the L$_3$ XMCD is too small.

Bekenov \textit{et al.} have calculated the XAS/XMCD spectra of CTS \textit{ab initio} using the spin polarized relativistic linear-muffin-tin-orbital (SPR LMTO) method.\cite{Bekenov05} Their simulations do not reproduce the double-peak structures and are rather similar to our SPRKKR spectra.

In FEFF9, the SPRKKR spectrum can be principally reproduced when the ground state density is used. Instead, if the density in the presence of a screened core hole is calculated, we find a structure that is very similar to the experimental spectrum. Because the self-consistency algorithm of FEFF9 is only accurate within 1 eV in its determination of the Fermi energy, one can use a small energy shift for fitting, thereby moving E$_F$ within the density of states (DOS). With a shift of -0.2\,eV we obtained the spectrum shown in Fig. \ref{fig:Co_comparison}. Obviously, both the double-peak structure of the white line as well as the small shoulder 4\,eV above the white line are reproduced. Also the double-peak structure of the XMCD signal is well reproduced. Notably, not only the shape of the spectrum is basically correct, but also the intensities match the experimental data very well. However, the double-peak splitting of the L$_3$ line is underestimated calculated as 1.3\,eV, compared to a measured splitting of 1.5\,eV.

Since FEFF9 is based on the local density approximation within the density functional theory---and thus relies on single-particle theory---it does not account for atomic multiplet effects, which naturally are many-body effects arising from wave-function coupling. Consequently, we conclude that the features observed in our experimental spectra do not arise from multiplet effects and electron localization. Instead, they are features arising from the excited state band structure due to the presence of a core-hole. This is consistent with the XMLD measurements discussed above, which indicate rather itinerant moments.

Our conclusion is further supported by the analysis given by Klaer \textit{et al.}, who found that the observed structures can not be explained by charge-transfer multiplet theory.\cite{Klaer09} They state that the splitting arises from a nearly pure Co e$_g$ state above E$_F$ giving rise to the first peak, and from a Co-Ti hybrid state of t$_{2g}$ character, which results in the second peak. Since the t$_{2g}$ states have more itinerant character, the core hole is more screened by the surrounding atoms, while the e$_g$ states are significantly lowered in energy. This core-hole correlation energy $\Delta E_C$ was assumed to be 0.5\,eV and confirmed by a measurement on Co$_2$TiSi. Neglecting the (only weak) energy dependence of the transition matrix elements, and using this core-hole correlation energy and spectral deconvolution, they finally found that the Fermi level of Co$_2$TiSn is at the edge of the minority valence band, i.e., Co$_2$TiSn would be on the verge of being a half-metal. With the same method, they found that Co$_2$MnSi has half-metallic character for the unoccupied density of states.

\begin{figure}
\begin{center}
\includegraphics[width=8cm]{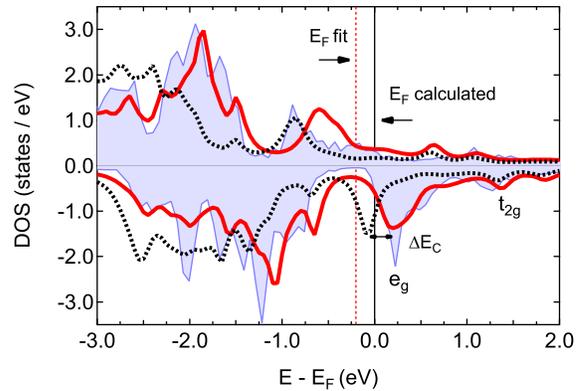}
\caption{Comparison of the calculated Co site projected \textit{d}DOS from SPRKKR (shaded blue area) and FEFF9 in the ground state (solid red line) and with an L$_3$ core hole (dotted black line).}
\label{fig:DOS_comparison}
\end{center}
\end{figure}

Using the FEFF9 calculations, we can invert this procedure. From \textit{ab initio} calculations we found the Fermi energy by fitting the experimental spectrum. Now we can use the same Fermi energy and investigate the ground state DOS calculated by FEFF9. The Co site projected \textit{d}DOS are shown together with the SPRKKR calculation in Fig. \ref{fig:DOS_comparison}. First, we shall note that the ground state DOS from FEFF9 and the SPRKKR calculation produce principally the same features, but FEFF9 underestimates the splitting between the bonding and the anti-bonding states. This is because of the spherical potential approximation and the use of the von Barth-Hedin exchange correlation potential. The unoccupied DOS is however in good agreement. Because of finite cluster size effects, the DOS from FEFF9 is broadened. The minority states gap can be identified just below the calculated Fermi level. When comparing the DOS in presence of the core hole to the ground state, we find that the curve is mainly shifted to lower energies by $\Delta E_C \approx 0.3$\,eV. In the unoccupied DOS, this is best seen for the minority e$_g$ peak, which shifts below the calculated E$_F$. Instead, the Co-Ti t$_{2g}$ peak at 1.4\,eV remains essentially unaltered. That is in remarkable agreement with the procedure given by Klaer \textit{et al.}. When the same Fermi level is applied to the ground state density as to the excited state density, we can conclude from our data that Co$_2$TiSn has half-metallic character with E$_F$ right below the minority valence band (see dotted energy level in Fig. \ref{fig:DOS_comparison}).

Finally, we shall discuss the limitations of our model. As mentioned above, the \textit{ab initio} calculation underestimates the double-peak splitting of the XAS by about 0.2\,eV. This introduces an uncertainty in the Fermi energy determination by spectral fitting of the order of the correction itself. With the currently available level of \textit{ab initio} theory this issue can not be resolved and it remains unclear if Co$_2$TiSn is a half-metallic ferromagnet. At least, a full potential treatment would be desirable, and spin self-consistency with more advanced exchange correlation functionals may help to resolve problems with the exchange splitting. On the other hand, the SPRKKR calculation finds the t$_{2g}$ peak at slightly lower energy than FEFF9. Thus it is possible that a more accurate calculation of the XAS requires approaches going beyond DFT.

\section{Conclusions}
We have grown thin films of the Heusler compound Co$_2$TiSn by DC magnetron co-sputtering. Structural investigations revealed highly ordered, fully epitaxial growth of Co$_2$TiSn thin films on MgO (001) substrates at growth temperatures above 600$^\circ$C.

A low residual resistivity supports the conclusion of well ordered films. The resistivity has a pronounced cusp-type anomaly at $T_C$. A large magnetoresistance has been observed and can be explained in terms of spin fluctuations.

From the XMCD measurements we find a total magnetization of $1.98\pm0.05\,\mu_B$\,/\,f.u., where the uncertainty arises from the unknown systematic error in the estimate of the Ti spin moment; the reduced average saturation magnetization of the best film (T$_S = 700^\circ$C, $m = 1.6(1)\,\mu_B$\,/\,f.u.) can be easily explained by an oxidized bottom interface layer of 3\,nm thickness.

The results for the element specific spin and orbital magnetic moments are in quantitative agreement with \textit{ab initio} band structure theory. The fine structures observed for the Co L$_{3,2}$ edges were explained by direct calculations of the XAS using FEFF9. Inclusion of the core-hole potential was found to reproduce the split white lines. A strong electron localization can be ruled out, in agreement with XMLD results. However, it remains unclear whether Co$_2$TiSn is a half-metallic ferromagnet or not.

\begin{acknowledgments}
The authors gratefully acknowledge financial support by the Deutsche Forschungsgemeinschaft (DFG)  and the Bundesministerium f\"ur Bildung und Forschung (BMBF). They thank for the opportunity to work at BL 6.3.1 and BL 4.0.2 of the Advanced Light Source, Berkeley, USA, which is supported by the Director, Office of Science, Office of Basic Energy Sciences, of the U.S. Department of Energy under Contract No. DE-AC02-05CH11231. They further acknowledge the work of Prof. Rehr and his group for developing and providing the FEFF9 code, as well as the work of Prof. Ebert and his group for developing and providing the \textit{Munich} SPRKKR package. 
\end{acknowledgments}

\appendix*

\section{}
To calculate spin and orbital moments from the absorption spectra we applied sum-rule analysis \cite{Chen95} to the XMCD spectra. 
Thereby we define the integrals $p$, $q$ and $r$ as
\begin{eqnarray*}
p&=&\int_{L_3}(I^{+}-I^{-})dE   \\
q&=&\int_{L_3+L_2}(I^{+}-I^{-})dE \\
r&=& \int_{L_3+L_2}\left( \frac{I^{+}+I^{-}}{2}-S \right) dE    \\
\end{eqnarray*}
We used a no-free-parameter two-step-like background function $S$ with thresholds set to the points of inflection on the low energy side of the $L_3$ and $L_2$ white lines and step heights of 2/3 ($L_3$) and 1/3 ($L_2$) of the average absorption coefficient in the post-edge region ("post-edge jump height $\eta$") here.
Sufficiently far away from the absorption edges and for photon energies above 30eV interactions among the atoms in the samples can be neglected \cite{XDB} and the post-edge jump height $\eta$ is proportional to $\sum_i X_i\sigma _{ai}$, where $X_i$ is the relative concentration of atom $i$ in the sample and $\sigma_{ai}$ is its total atomic absorption cross section.
As pointed out by  St\"ohr \cite{Stoehr95}, the number of unoccupied 3d states $N_h$ is proportional to the integral $r$ via $r=CN_h\eta$. The constant $C$ depends on the transition matrix elements connecting the core and valence states involved in the $2p$ to $3d$ transitions and has been analyzed by Scherz for different $3d$ transition metals ($C^{Ti}=5.4$eV, $C^{Co} = 7.8$eV).\cite{ScherzPhD} We used the integral $r$ to determine $N_h$ for the different samples. While neglecting the spin magnetic dipole term $<T_Z>$ in the XMCD sum rules, the spin and orbital magnetic moments $m_\text{spin}$ and $m_\text{orb}$ and their ratio are then given as
\begin{equation}
m_{orb}=-\frac{1}{P_{h\nu}\cos{\theta}}\frac{4q }{6C\eta}
\label{eq3a}
\end{equation}
\begin{equation}
m_{spin}=-\frac{1}{P_{h\nu}\cos{\theta}}\frac{(6p-4q)}{2C\eta}
\label{eq3b}
\end{equation}
\begin{equation}
\frac{m_\text{orb}}{m_\text{spin}}=\frac{2q}{9p-6q}
\label{eq3c}
\end{equation}
with the elliptical polarization degree $P_{h\nu}$  and the angle $\theta$ between magnetization and x-ray beam direction.

\end{document}